\documentclass[a4paper,12pt]{article}
\usepackage{verbatim} 
\usepackage{jheppub} 

\usepackage[T1]{fontenc} 

\usepackage[percent]{overpic}
\usepackage{wrapfig}
\usepackage{tabu}
\usepackage{diagbox}

\usepackage{graphicx}
\usepackage{tikz}
\usepackage{import}
\usepackage{accents}
\usepackage{mathrsfs,amsmath,amssymb,slashed}
\usepackage{multirow,multicol}
\usepackage[percent]{overpic}
\usepackage{slashed}
\newcommand{\pii}{\textbf{p}}
\newcommand{\x}{\text{x}}
\usepackage{wrapfig}
\usepackage{tabu}
\usepackage{diagbox}
\usepackage{mathrsfs,amsmath,amssymb,amsthm,amsfonts,tikz,graphicx,accents,hyperref,color}
\usepackage{leftidx}
\usepackage{import}
\usetikzlibrary{decorations.pathmorphing}
\DeclareFontFamily{OT1}{rsfs}{}

\DeclareFontShape{OT1}{rsfs}{m}{n}{ <-7> rsfs5 <7-10> rsfs7 <10->rsfs10}{} 

\DeclareMathAlphabet{\mycal}{OT1}{rsfs}{m}{n}

\newcommand{\Laa}{\Lambda}

\newcommand{\mm}{\mathbf{m}}

\newcommand{\B}{\mathcal{B}}

\newcommand{\Hh}{\mathcal{H}}

\newcommand{\ord}[1]{\mathcal{O}(#1)}
\newcommand{\laa}{\lambda}

\newcommand{\di}{\text{d}}

\newcommand{\A}{\mathcal{A}}

\newcommand{\F}{\mathcal{F}}
\newcommand{\p}{\prime}
\newcommand{\en}{\mathbf{n}}

\newcommand{\eq}[2]{\begin{equation}
#1\label{#2}
\end{equation}}
\newcommand{\nn}{\nonumber}

\newcommand{\be}{\begin{equation}}
\newcommand{\ee}{\end{equation}}

\DeclareMathOperator{\extdm}{d}
\newcommand{\extd}{\extdm \!}

\newcommand{\cF}{\mathcal{F}}

\newtheorem{defi}{Definition}[]
\newtheorem{pro}{Proposition}[]

\title{\centerline{\boldmath {String Memory Effect}}}
\author[]{Hamid  Afshar,}
\author[]{Erfan Esmaeili,}
\author[]{M. M. Sheikh-Jabbari}
\affiliation[]{\it School of Physics, Institute for Research in Fundamental
Sciences (IPM),\\ P.O.Box 19395-5531, Tehran, Iran}
\emailAdd{afshar@ipm.ir}
\emailAdd{erfanili@ipm.ir}
\emailAdd{jabbari@theory.ipm.ac.ir}

\preprint{IPM/P-2018/080}

\abstract{
In systems with local gauge symmetries, the memory effect corresponds to traces inscribed on a suitable probe when a pure gauge configuration at infinite past dynamically evolves  to another pure gauge configuration at  infinite future. In this work, we study the memory effect of 2-form gauge fields which is probed by strings. We discuss the ``string memory effect'' for closed and open strings at classical and quantum levels. The closed string memory is encoded in the internal excited modes of the string, and in the open string case, it is encoded in the relative position of the two endpoints and the noncommutativity parameter associated with the D-brane where the open string endpoints are attached. We also discuss  2-form memory with D-brane probes using boundary state formulation and, the relation between string memory and 2-form soft charges analyzed in \cite{Afshar:2018apx}.
}
\begin{document}
\maketitle

\section{Introduction}

Conserved charges 
associated with global symmetries through the celebrated Noether theorem have been the cornerstone of analyzing physical processes, especially in  scatterings. Some of these physical processes, while  can be understood as scatterings of matter fields (probes) off massless gauge fields (e.g. photons or gravitons), may also be analyzed as  \emph{memory effects}. In these specific cases, the initial and the final state of the gauge field or the metric differ by a pure gauge or coordinate transformation, while  there is a non-zero field strength in the middle of the process, due to the radiation.  The memory effect is the imprint of this evolution of the gauge field from far past to  far future  on an appropriate probe. These processes involve very low energy \emph{soft photons/gravitons},
and one may hence expect them to be closely related to  other IR effects in gauge theories, like (Weinberg's) soft theorems. One may wonder whether  these IR effects can also be analyzed using the notion of conserved charges.  We are therefore, dealing with the ``IR triangle'' \cite{Strominger:2013lka,Strominger:2013jfa,He:2014cra,Strominger:2014pwa,Strominger:2017zoo} which relates soft theorems, asymptotic charges and memory effects.

Asymptotic symmetries involve a specific set of gauge transformations remaining after certain gauge fixing and preserving  certain boundary conditions. Asymptotic charges associated with these symmetries can differentiate the initial and final photon/graviton states. 
 Therefore, the change in the value of the charge can be stored in the probe as memory effect.

Depending on the quantity imprinted on the probe we may have different types of memory effects. Historically, the first memory effect was discussed in a gravitational context where passage of gravity waves was imprinted in displacements of a geodesic congruence \cite{Christodoulou:1991cr,Zel'dovic}. This memory effect in the language of soft charges is associated with supertranslations. There are also two other kinds of gravitational memories: spin and refraction memory effects, respectively associated with superrotations \cite{Pasterski:2015tva} and superboosts \cite{Compere:2018ylh}; see also \cite{Laddha:2018vbn,OLoughlin:2018ebk,Zhang:2018srn}.
In electrodynamics the leading memory effect \cite{Bieri:2013hqa} is the shift of the velocity of a charged point particle under the 1-form gauge field radiation (the kick memory effect) --- see also \cite{Winicour:2014ska,Susskind:2015hpa,Pasterski:2015zua,Mao:2017axa}. 

Antisymmetric gauge fields which are ubiquitous in theories of supergravity and string theory also come with a gauge symmetry; a $(p+1)$-form (with a $(p+2)$-form field strength) enjoys a gauge symmetry generated by  $p$-forms. Unlike the electromagnetic case, objects which are charged under  $(p+1)$-forms are spatially extended {$p$-branes} with a $(p+1)$-dimensional worldvolume. 
Generically, $p$-branes (with $p>0$) have internal degrees of freedom which can be degenerate, so that the transition between these internal modes has no energy cost. In particular, there could be transitions between these modes by the exchange of \emph{soft $(p+1)$-form} modes. This latter yields a new type of memory effect, the ``internal memory effect'', which is associated with the change in internal excitations of an extended object, like a $p$-brane. The internal memory effect may be put in contrast with other so-far discussed memories, the ``external memory'', which are associated with the change in a spacetime property of the probe like position, momentum or spin. 

In this work we intend to study such an internal memory effect by focusing on the $2$-form theory (the $p=1$ case) which is naturally coupled to strings. This specific example, while carries aspects of the higher form cases, has its own unique features. In particular, due to the conformal symmetry of the worldsheet theory, we completely know the spectrum of free strings and have  much better control on the analysis. Moreover, the $2$-form gauge field background as well as the fundamental string probes, are both contents of the string theory setup and prepare a framework to study ``string memory effect''. We also connect the string memory effect to those soft charges and asymptotic symmetries recently studied \cite{Afshar:2018apx} (see also \cite{ Campiglia:2018see, Francia:2018jtb}) to construct one edge of the $2$-form soft triangle. 

In order to study the string memory effect, we revisit the problem of strings in a slowly varying Neveu-Schwarz--Neveu-Schwarz (NSNS) 2-form $\B$-field background. It is known that the $\B$-field affects closed and open strings in different ways and hence we consider these two cases separately. As we show, for the closed string case the string memory effect is encoded in the  transition of the massless states of the closed string into each other.  Thus, in the string memory effect the probes themselves can be massless states, in contrast with  other usual memories in which the probe is a massive state.\footnote{Note that in the internal string memory effect the probe can also be one of the massive modes of string excitations.}

 In the open string, the constant part of the $\B$-field affects string dynamics through the boundary conditions. It is known that the endpoints of open strings attached to a D-brane in the $\B$-field background parameterize a noncommutative space \cite{Ardalan:1999av, SheikhJabbari:1999vm, Chu:1998qz,Chu:1999gi, Seiberg:1999vs}. The open string memory effect is then encoded in the change in the noncommutativity of the open string endpoints. As we will discuss, this memory can hence be observed through the effective noncommutative field theory residing on the brane where the open string endpoints attach. Alternatively, one may use the D-branes as probes of the background time-varying $\B$-field, using boundary state formulation for D-branes \cite{DiVecchia:1997vef, Arfaei:1999jt}. In this system, the change in the $\B$-field is encoded in the mass density or other Ramond-Ramond (RR) charges of the D-brane probe. The similar problem of strings in a gravitational background pulse was considered in \cite{Bachas:2002jg} and the possibility of permanent shift in brane separation was discussed, which resembles point particle gravitational memory effects. 
 
 \paragraph{Organization of the paper.} In section \ref{sec:2}, to set the stage we review the memory effect especially in the case of electromagnetism. We discuss the electromagnetic kick memory effect at classical and quantum levels. As we will review, the memory on quantum probes can be expressed by an $S$-matrix. The quantum memory effect is what is relevant to our string theory analysis. In section \ref{sec:3}, we study the closed string memory effect and discuss how the passage of a $\B$-field wave causes a shift in the internal mode of the closed string, without affecting its center of mass motion. In section \ref{sec:4}, we analyze the open string memory effect caused by the passage of an NSNS $\B$-field wave packet. This memory is imprinted in the noncommutativity parameter associated with the open string endpoints. In the last section, we discuss physical implications and possible extensions of our analysis here, as well as other open questions for future analysis. In an appendix, we have discussed the absence of electromagnetic memory on a harmonic oscillator probe.

\subsection*{Notation and conventions:}
\textbf{Spacetime coordinates.}
We are mainly interested in an experiment performed at the  radiation zone of a scattering process which involves the emission of $2$-form soft photons, see \cite{Afshar:2018apx} for more details. In the radiation zone, if the size of the measurement apparatus is much smaller than the radius of the sphere surrounding the scattering region, wavefronts can be well approximated as plane waves.  In the vicinity of a point on the sphere the line element is
\begin{equation}\label{lineelement}
    ds^2=-du^2-2dudr+\delta^{ij}dX^idX^j\,,
\end{equation}
where $u=t-r$ is the retarded time, $X^i$ are local Cartesian coordinates in vicinity of a point $p$ on the ($D-2$)-sphere  with $i=1,2,\cdots, D-2$ and $r$ is the radial outward pointing coordinate.  For transverse Cartesian coordinates we use $i,j,m,l,$ etc.  
Finally Greek indices $\mu,\nu,\alpha$ etc. go on all coordinates including $u$.

The fall-off behavior of fields are most often expressed in spherical coordinates $\theta^A$  with metric $r^2q_{AB}d\theta^Ad\theta^B$, which covers the whole sphere. It is always presumed that, if the large-$r$ behaviour of  (say) a 1-form in spherical coordinates is $T_{A}\sim \ord{r^{-n}}$, then $\partial_{B}T_A\sim \ord{r^{-n}}$. The implicit premise here is that $T_A$ does not vary too fast on the sphere, so its value and its derivative have the same magnitude. This will be our working premise too.\footnote{This means that we are neglecting large $\ell$ components of the field in a spherical harmonics expansion.} One should, however, note  that when working in Cartesian coordinates on a small patch of the sphere, one has
\begin{align}
    T_A\sim \ord{r^{-n}}\,&\quad\rightarrow \quad T_i=e^A_iT_A\sim \ord{r^{-n-1}}\,,\\
        \partial_{B}T_A\sim \ord{r^{-n}}\,&\quad\rightarrow \quad \partial_jT_i=e^B_je^A_i\partial_{B}T_A\sim \ord{r^{-n-2}}\,,
\end{align}
where $e_i^A$ are changing the Cartesian to spherical basis. That is, each Cartesian derivative lowers the $r$-power by one unit.

\paragraph{Worldsheet.} Worldsheet coordinates are $(\tau,\sigma)$ and derivatives with respect to them are denoted by dot and prime respectively. For closed strings $\sigma\in[0,2\pi]$ while for open strings $\sigma\in [0,\pi]$. The string length is $\ell_s\equiv \sqrt{4\pi\alpha^\p}$. The left/right worldsheet derivatives are defined as
\begin{equation}
  \partial_+=\partial_\tau+\partial_\sigma \qquad
     \partial_-=\partial_\tau-\partial_\sigma \,.
\end{equation}
We will work in light-cone gauge $u=\ell_s\tau$ throughout. Embedding coordinates of the string into the target space will be denoted by $X^\mu(\tau, \sigma)$, which  are divided into $u, r(u,\sigma), X^i(u,\sigma)$ in  light-cone gauge. When dealing with D$_p$-branes we use $X^I, \ I=0,\cdots, p$ for directions along the brane.

\paragraph{Spacetime fields.} The fields in spacetime are denoted by curly letters, while for the coefficients of asymptotic expansions, normal fonts are used. The 1-form electromagnetic gauge field is denoted by $\A$ and its field strength by $\F=d\A$. The 2-form gauge field is shown by $\B$ and its field strength by $\Hh=d\B$.
 

\section{Memory effect: review and remarks}\label{sec:2}

In this section we discuss three issues. First, we give a definition for memory effect with the focus on Hamiltonian formulation. Second, we revisit electromagnetic memory effect on a free particle. Third, we discuss memory effect for the quantized version of the same system.

Consider a physical system with classical phase space coordinates $(q_n,p_n)$ and the Hamiltonian $H$. Let the system interact with the environment through a \emph{gauge interaction} with a gauge potential that we schematically denote by  $\A$. In typical examples, the classical equations of motion are gauge invariant, depending exclusively on a  gauge invariant field strength $\F$.  For instance, the form of Lorentz force in electromagnetism is proportional to $\F_{\mu\nu}\dot X^\nu$. Also, observable quantities $\mathcal{O}[q,p;\A]$ in systems of physical interest are always gauge invariant. In other words, $\mathcal{O}[q,p;\A]=\mathcal{O}[q,p;\tilde{\A}]$ where $\A$ and $\tilde{\A}$ are related by a gauge transformation generated by $\Lambda$. 
Since gauge transformations of the external field leave all observables of the system invariant, one expects them to be equivalent to 
 a canonical transformation  $U[q,p;\Laa,\A]$ on phase space $(q_n,p_n)$. In particular,  if the field strength $\F$ is vanishing, one can find a canonical transformation $U$ to set $\A=0$ (up to possible topological obstructions). We will find such transformations in the examples we analyze. 
 
We will assume that the external field $\A$ is only a function of the retarded time; justified physically as follows. 
If the radius of the  sphere on which the probes (point particle, strings, etc.) are located is large enough compared to the characteristic length of the probe (i.e. $x_0$,
$r\gg x_0$), 
 then $\A$ is almost constant in the vicinity of the probe. This assumption fails if $\A$ is highly oscillating on the celestial sphere 
which will be disregarded in current calculation.

\begin{pro}\label{canonic}
Let a classical system $(p_n,q_n;H)$ interact through an external gauge potential $\A$ with time-dependent field strength $\F(u)$ which is  vanishing at early and late times:
\eq{
\lim_{u\to\pm\infty}\F(u)=0\,,
}{}
so the system evolves with free Hamiltonian in the limit $u\to\pm\infty$.\footnote{This is typically the case when $\F(u)$ is a radiative field which is strong only in a finite duration.} Defining the pure gauge configurations,
\begin{align}
    \A^+\equiv \A(u=+\infty),\qquad
    \A^-\equiv\A(u=-\infty),
\end{align}
the system at very early and late times is governed by free Hamiltonians $H[\A^+]$ and $H[\A^-]$. These two Hamiltonians  are related by a canonical transformation $U[q,p;\A^+,\A^-]$.
\end{pro} The time evolution of the system is  \emph{free} at large $|u|$.
As a result, the phase space and the Hamiltonian at $u\to\pm\infty$ are related by a canonical transformation $U[q,p;\A_+,\A_-]$ and free solutions are mapped to free solutions by the same operator.

In the above setting, we propose the following definition for the memory effect:
\begin{defi}[Memory effect]
Let $\mathcal{O}[q,p;\A]$ be an observable of the system described above, for which the following quantities are well-defined:
\begin{align}
    \mathcal{O}^+=\lim_{u\to+\infty}\mathcal{O}[q,p;\A],\qquad 
    \mathcal{O}^-=\lim_{u\to-\infty}\mathcal{O}[q,p;\A].
\end{align}
If $\mathcal{O}^+$ and $\mathcal{O}^-$ are \underline{not related} to each other by the canonical transformation defined in proposition \ref{canonic}, we say that gauge field $\A$ has induced a memory effect on the system.
\end{defi}

\subsection{Classical treatment}
Although most of the work on memory effect has been done for classical systems, one can figure out the effect on the state of a quantum system. We first consider the classical approach, and analyze memory effect in Hamiltonian formulation. This latter sets the stage for formulating memory effect at quantum level.
\paragraph{Electromagnetic Memory Effect.}

Before analyzing the memory effect exerted by a 2-form gauge field  on a string probe, we first review  the memory effect in the case of electromagnetic 1-form gauge field coupled to a point particle of charge $q$ and mass $m$. One may analyze the problem in action or Hamiltonian formulations. Let us start with the action and equations of motion:
\begin{equation}
    S=\int d\tau \,\eta\left(\frac{1}{2}\eta^{-2}\partial_\tau X^\mu\partial_\tau X^\nu g_{\mu\nu}+q\,\eta^{-1}\partial_\tau X^\mu\A_\mu-\frac{1}{2}m^2
    \right),
\end{equation}
where $\eta(\tau)$ is the einbein of the metric on the worldline $ds^2=\eta^2(\tau)d\tau^2$. We take $X^\mu=(u,r,X^i)$ as Minkowski coordinates  defined in \eqref{lineelement}.\footnote{In order to capture the physics of radiation at null infinity we expand all quantities in inverse powers of $r$ in \eqref{lineelement} with expansion coefficients which are functions of retarded time $u$ and  Cartesian coordinates $X^i$.
} This action is gauge invariant up to a boundary term and we can fix the radial gauge $\A_r=0$. To fix the reparametrization freedom we use the light-cone gauge which sets $u$ as the clock, i.e. $u=\tau$. The action is then
\begin{equation}\label{worldline}
    S=\int d\tau \left(\frac{1}{2\eta}(-1-2\dot{r}+\dot{X}^i\dot{X}^j\delta_{ij}) +q\left(\dot{X}^i\A_i+\A_u\right)-\frac{1}{2}\eta\, m^2
    \right).
\end{equation}
Assuming that the gauge field components $\A$  have very mild $r$ dependence (according to fall-off behavior given below), the equation of motion for $r$ yields $\eta\approx const$ and we must choose this constant equal to $1/m$ to be consistent with non-relativistic limit, taken below.\footnote{\label{constraint-footnote}Variation of the action \eqref{worldline} with respect to $\eta$ gives $\dot{X}^\mu\dot{X}_\mu+\eta^2m^2=0$ as a constraint.}
Equations of motion for $X^i$ then reads as
\begin{align}\label{EMeom}
\ddot {X}^i=\frac{q}{m} (\cF^i_{\ j} \dot X^j+\cF^i_{\ u})\,. 
\end{align}
We now focus on cases where the particle is coupled to the radiation photon field through $\A_i$ with  the  boundary conditions at large $r$,
\begin{align}\label{A-falloff}
    \A_i\sim\ord{1/r}\,,\qquad \A_u\sim\ord{1/r}\,,\qquad \A_r=0\,.
\end{align}
The field strength component $\F_{ui}$ is at order $r^{-1}$, while the other components of the field strength $\F_{ij}$ and $\F_{ur}$ are subleading. 
Integrating \eqref{EMeom} along the retarded time $u$, we are left with an electric \emph{kick memory effect},
\begin{align}\label{kickmemory}
    \Delta \dot{X}^i=\dot{X}^i(u=\infty)-\dot{X}^i(u=-\infty)&\simeq \frac{q}{m}\int_{-\infty}^{\infty}\cF^i{}_u\extd u\nn\\
    &=-\frac{q}{m r}\left(A_i(u\to\infty)-A_i(u\to-\infty)\right)\,,
\end{align}
in which $A$ is the leading term in the asymptotic falloff \eqref{A-falloff} and terms of $\ord{r^{-2}}$ are ignored.

In order to apply and use the results of the proposition \ref{canonic}, we  present these results in the Hamiltonian formulation. 
From \eqref{worldline} the canonical momentum $\check{P}_i$ and $\check{P}_r$ conjugate to the position coordinates $X^i$ and $r$ are,
\begin{equation}
    \check{P}_i=\frac{1}{\eta}\dot{X}_i+q\,\A_i\,,\qquad \check{P}_r=-\frac{1}{\eta}\,.
\end{equation}
The Hamiltonian is
\eq{
H=-\frac{m^2+\check{P}_r^2+(\check{P}_i-q\,\A_i)^2}{2\check{P}_r}-q\,\A_u\,.
}{}
The constraint equation (\emph{c.f.} footnote \ref{constraint-footnote}), $\eta^2 m^2-1-2\dot{r}-(\check{P}_i-q\A_i)^2=0$ associated with  the reparametrization invariance of the probe action can be solved for $\eta$. Thus, for a  non-relativistic particle, deviation of $\eta$ from $1/m$ is small, suggestive of  defining a shifted radial momentum $P_r\equiv \check{P}_r+m\ll1$.  Finally, the Hamiltonian for a non-relativistic charged particle of rest mass $m$ moving in  the  light-cone gauge $\tau=u$ for the particle and in the radial gauge $\A_r=0$ for the  gauge field, takes the familiar form
\eq{
H=m+\frac{P^2_r+(\check{P}_i-q\A_i)^2}{2m}-q\,\A_u\,,\qquad i=1,2\,.
}{classical hamilton}
We drop the rest mass constant term $m$ from now on.
Clearly, the Hamiltonian \eqref{classical hamilton} takes different forms at $u\to\pm\infty$.  
Recalling the falloff behavior of the gauge field at large $r$ \eqref{A-falloff}, the particle is ``free'' at order $r^{-1}$ and one expects   early and late Hamiltonians to be related by a canonical transformation. To show this,
consider new dynamical variables $(P,X)$ related to old variables $(\check{P},X)$ by the canonical transformation on momenta
\begin{equation}\label{free-elect-canonic}
    P_r=\check{P}_r+m\,,\qquad P_i=\check{P}_i-q\,\A_i\,,
\end{equation}
while the coordinates remain unaltered. The new Hamiltonian $K$ is
\begin{equation}
K=\frac{P_r^2+P_i^2}{2m}+q\,X^i\partial_u\A_i\,,\qquad i=1,2\,.
\end{equation}
To verify this statement, we note that new and old variables in the large $r$ limit are related as
\begin{equation}\label{genrel}
    \dot{X}^a\check{P}_a-H=\dot{X}^aP_a-K+\frac{d{G}}{du}\,,\qquad a=1,2,3,
\end{equation}
 and the generating function\footnote{The function $G$ as appears in \eqref{genrel} does not generate the transformation. Different kinds of generating functions are obtained only by adding certain combinations of old and new coordinates and momenta (like $XP$) to $G$.} for the transformation is 
\begin{equation}G= qx^i \A_i+q\int_{u_0}^u\A_u(u^\p)du^\p.
\end{equation}
The integral term in $G$ does not appear in transformation relations \eqref{free-elect-canonic}; it becomes subleading in $r$ when Cartesian derivatives are performed.

The new Hamiltonian $K$ has the same form at $u\to\pm\infty$ since the electromagnetic radiation $\F_{ui}=\partial_u\A_i+\ord{r^{-2}}$ by assumption vanishes at both temporal limits. We are now ready to find an observable in the new coordinate system which has different asymptotic values at early and late times. The asymptotic form of equations of motion in the new basis are
 \eq{
 \dot{X}^a=[X^a,K]=\frac{P_a}{m}\,,\qquad \dot{P}_i=[P_i,K]=-q\,\dot{\A}_i\,,\qquad\dot{P}_r=0\,.
 }{}

\begin{pro}[Memory effect]
Electromagnetic  Memory effect on a free charged non-relativistic particle is the difference between late and early transverse momenta:
\begin{equation}\label{kickmemory2}
\Delta P_i=-q\,\Delta \A_i\,,
\end{equation}
where for a generic variable $V$, $\Delta V$ measures the difference between the late and early values, 
\eq{\Delta V\equiv V(u=+\infty)-V(u=-\infty)\,.}{Deltadef} 
\end{pro}
Eq. \eqref{kickmemory2} is the analogue of \eqref{kickmemory} in the Hamiltonian formulation. 
The above example provides a smooth transition to a quantum treatment of the  memory effect.

\subsection{Quantum treatment}

The analysis of previous section in Hamiltonian formulation can be quantized in a straightforward way, leading to electromagnetic memory effect on the momentum shift of quantum charged particles on the corresponding Hilbert space. The canonical transformation which brings the Hamiltonian to the convenient form becomes a unitary transformation on the  Hilbert space.  Consider the quantized Hamiltonian \eqref{classical hamilton}
\eq{\hat{H}= \frac{\hat{P}^2_r+\left(\hat{P}_i-q\A_i(\hat{X})\right)^2}{2m}-q \A_u(\hat{X})\,,
}{Q-Hamiltonian}
and the unitary operator\footnote{Hatted symbols refer to quantum operators.}
\begin{equation}
    \hat{U}=\exp{\left(-iq \hat{X}^i\A_i(\hat{X})-iq\int_{u_0}^{u}du^\p \A_u(\hat{X})\right)}\,.
\end{equation}
Neglecting subleading terms in $\hat{r}$, $\A_i(\hat{X})=\A_i(u)$ and $\A_u(\hat{X})=\A_u(u)$, so the unitary transformation will asymptotically act as follows:
\begin{align}
   & U\hat{X}_aU^\dagger =\hat{X}_a\,,\\
   &U\hat{P}_iU^\dagger =\hat{P}_i+q\,\A_i(u)\,,\\
 &U\hat{P}_rU^\dagger =\hat{P}_r\,,
\end{align}
while transforming the Hamiltonian to 
\begin{align}\label{Kamilton particle}
    \hat{K}&=\hat{U}\hat{H}\hat{U}^\dagger+i\frac{\extd{\hat U}}{\extd u}U^\dagger
=\frac{\hat{P}_i\hat{P}_i+\hat{P}_r^2}{2m}+q\,\hat{X}^{ i}\frac{\extd{\A}_i(u)}{\extd u}\,.
\end{align}
Time evolution of a state vector $|\Psi\rangle$ in this canonical frame is given by 
\eq{
|\Psi(u)\rangle=e^{-i(u-u_0)\hat{K}}|\Psi(u_0)\rangle\,.
}{arg2}

In the $u\to\pm\infty$ limit, the \emph{in}/\emph{out} Hilbert spaces can be written in the basis of eigenstates of the free Hamiltonian ($K$ at $q=0$). We want to find the \emph{S-matrix} operator $\hat{S}:\mathcal{H}_{in}\to\mathcal{H}_{out}$ which quantifies the overlap of \emph{in} and \emph{out} states.
\begin{pro}
 If a quantum particle with Hamiltonian \eqref{Kamilton particle} is prepared in an energy eigenstate $|in\rangle=|\en\rangle$ at $u\to-\infty$, then its state at $u\to+\infty$ is given by 
\begin{equation}\label{in-out-S}
    |out\rangle=\hat{S}|in\rangle\,,\qquad \hat{S}=\exp{\left(-iq\,\hat{X}_{\text{int}}^i\,\Delta{\A}_i\right)}\,,
\end{equation}
where $\Delta{\A}_i$ is defined  in \eqref{Deltadef}, and $\hat{X}_{\text{int}}=\hat{X}-\hat{P}u/m$ is the interaction-picture position operator. The quantum memory effect can also be expressed as an operator equation in Heisenberg picture
\begin{equation}
    \hat{S}^{-1}\hat{P}_i\,\hat{S}=-q\,\Delta{\A}_i\,.
\end{equation}
\end{pro}
To show \eqref{in-out-S}, we start with the more  precise statement of adiabatic evolution by  solving the problem for a particle in a large 3D box of dimension $L$. The free spectrum is
\eq{
\frac{\hat{P}_a\hat{P}_a}{2m}|\en\rangle=E_\en|\en\rangle,\qquad\qquad E_\en=\frac{\pi^2}{2mL^2} \en\cdot\en
}{}
Let $|\en\rangle$ be the interaction picture state, so that its time evolution be given by the interaction term in the Hamiltonian:
\eq{
i\frac{d}{du}|\en\rangle=q\,e^{-iK_0u}\hat{X}_i\dot{\A}_ie^{iK_0u}|\en\rangle=q(\hat{X}-\hat{P}u/m)^i\dot{\A}_i|\en\rangle\,\equiv q\hat{X}^i_{\text{int}}\dot{\A}_i|\en\rangle\,.
}{}
The solution is
\begin{equation}
    |\en(u)\rangle=\exp\left(-iq\,\hat{X}_{\text{int}}^i\,\A_i\Big|^u_{u_0}\right)|\en(u_0)\rangle\,,
\end{equation}
being in particular true for $in/out$ states.

As the above indicates, the momentum kick and memory is present for free particles. Nevertheless, one can  show that (see appendix \ref{HOSCharged}), for bounded particles (e.g. in harmonic oscillator) the memory for both classical and quantum systems, is averaged out in time.

\subsection{Memory effect and conserved charges}
In this subsection we relate the `memory effect' to a change in (soft) charges associated to non-trivial large gauge transformation from space-like infinity $i_0\equiv \mathcal{I}_-^+$ to future time-like infinity $i_+\equiv \mathcal{I}_+^+$ as two endpoints of future null infinity\footnote{More precisely: spacelike infinity ($i^0$) is the destination of all spacelike radial curves. $\mathcal{I}^+_-$ is a boundary of $i^0$ where the spacelike curves tend to be null. The same statement holds for timelike infinity ($i^+$) and $\mathcal{I}^+_+$.}. 
Conserved charges associated with large gauge transformations can be obtained as integrals of the time component of the  
current $J^\mu$, which can in turn be (locally) written as divergence of a two-form $J^\nu=\nabla_\mu {\mathcal J}^{\mu\nu}$, on a spacelike Cauchy surface $\Sigma$. In the Bondi-slicing of flat spacetime  we have
\begin{align}
    Q=\int_{i^0}\extd\Omega_{d-2}\,{\mathcal J}^{ru}\,.
\end{align}
Once the boundary conditions are imposed appropriately, this boundary charge is finite, conserved and integrable. In the case of electromagnetism the current associated with the
gauge parameter $\lambda$ is ${\mathcal J}^{\mu\nu}[\Lambda]=\lambda \,\F^{\nu\mu}$.
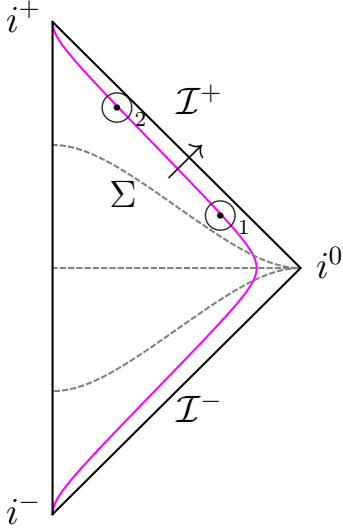
\begin{figure}[t]
 \centering
 \begin{overpic}[width=0.2\textwidth,tics=1]{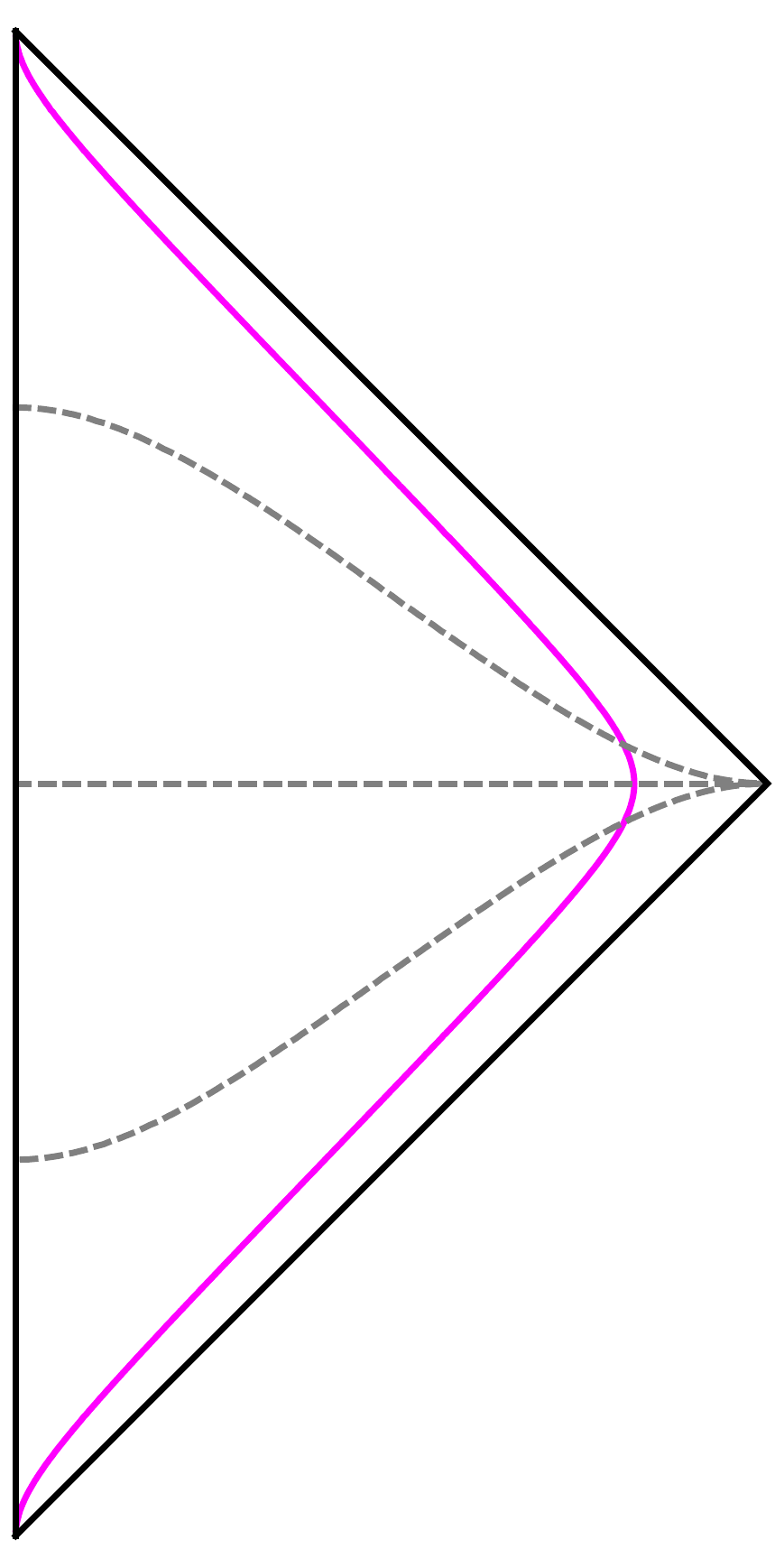}
 \put(23,69) {\large$\nearrow$}
 \put(12,62)  {\large$\Sigma$}
 \put(10,79)  {\Large$\odot_{\text{\scriptsize2}}$}
 \put(30,58){\Large$\odot_{\text{\scriptsize1}}$}
 \put (25,20) {\large$\mathcal{I}^-$}
  \put (25,80) {\large$\mathcal{I}^+$}
   \put (52,48) {\large$i^0$}
 \put (-8,0) {\large$\displaystyle i^-$}
 \put (-8,96) {\large$\displaystyle i^+$}
\end{overpic}
\caption{Observation of memory effect due to the out-going radiation by massive probes. The $\odot_{1,2}$ symbols show detection points, located at constant $r$. One could set the detection times earlier to observe the in-going memory effect. The dashed lines depict some Cauchy surfaces and as we see they all intersect at spatial infinity $i_0$. In the figure we have implicitly imposed the antipodal matching through $i^0=\mathcal{I}^+_-=\mathcal{I}^-_+$.}
    \label{Mink-de-Stter-comparison}
\end{figure}

 Soft charges at past $\mathcal{I}^+_-$ and future $\mathcal{I}^+_+$ of future null infinity are sensitive to the value of gauge field $\A(u=-\infty)$ and $\A(u=+\infty)$ there. Since memory effect depends on their difference $\Delta\A$, it will be related to the charge difference at two ends of null infinity. In the electromagnetic case, this difference is
 \eq{
 Q_\laa[\mathcal{I}^+_+]-Q_\laa[\mathcal{I}^+_-]=\int_{\mathcal{I}^+_+}\laa\, \F^{ur}-
 \int_{\mathcal{I}^+_-}\laa\, \F^{ur}=\int_{\mathcal{I}^+} \laa\,\partial_u \F^{ur}=\int_{S^2} \laa\,\mathcal{D}^i A_i\Big|_{u=-\infty}^{u=+\infty}
 }{softchargememory}
where in the last equality we used the field equation $\left(\extd\star{\cal F}\right)_u=0$ asymptotically. We also saw in \eqref{kickmemory} that electromagnetic memory effect is also given asymptotically by $\Delta A_i$ with definition as in \eqref{Deltadef}. This is the general pattern. 

The above equation may be read as follows: While ${\cal I}^+$ is not a Cauchy surface, $\Sigma^+={\cal I}^+\cup i^+$ is so and $i^0$ is the boundary of this and all other constant time Cauchy surfaces, \emph{cf.} dashed lines in Figure \ref{Mink-de-Stter-comparison}.\footnote{Note the ${\cal I}^+$ is defined at $r\to\infty$ with arbitrary $u$, and ${\cal I}^+_\pm$ is defined as $r\to\infty$ and then $u\to\pm\infty$.} Therefore, one expects that the soft charge $\int_{\Sigma^+} \lambda \F^{ur}$ to be conserved. However, massless particles can reach ${\cal I}^+$ and the massive probes to $i^+$. So, the change in the photon soft charge can be attributed to the change in the soft charge of the probe particle, a.k.a the memory.

\paragraph{Two-form soft charges.} In the case of 2-form theory, the value of the Noether charge associated to large gauge transformation at space-like infinity (past boundary of future null infinity) was derived in \cite{Afshar:2018apx} in six dimensions where it was shown that the conserved charges of the theory split into two separately conserved exact and coexact pieces associated to the exact and the coexact parts of the one-form gauge transformation parameter on the sphere $S^4$. In \cite{Afshar:2018apx}, the expression of the charge was derived in de Sitter slicing, while in the analysis for the memory effect we need the expressions on the null infinity  and the Bondi slicing. The details of the derivation of the charges in the latter slicing will be given elsewhere \cite{progress:EH} here we show the final result:
\begin{align}
Q^{\text{\tiny coexact}}[\lambda]=\int_{\mathcal{I}^+_-}\mathcal{J}^{ru}[\lambda]\,,\qquad Q^{\text{\tiny exact}}[\lambda]=\int_{\mathcal{I}^+_-}\mathcal{K}^{ru}[\lambda]\,,
\end{align}
where 
\begin{align}
{\mathcal J}^{\mu\nu}[\Lambda]=\Lambda_\alpha \Hh^{\alpha \nu \mu}
,\qquad 
{\mathcal K}^{\mu\nu}[\Lambda]=2(\di\Lambda)^{\alpha[\mu} {(\di{\mathcal C})^{\nu]}}_\alpha
\end{align}
are the corresponding currents associated to the coexact and exact charges. These expressions are for a two-form theory with Lagrangian ${\cal L}=\frac{1}{3!} {\cal H}^2,\ {\cal H}=\extd\B$ and $\Lambda$ is the one-form gauge parameter, associated with gauge transformation $\B\to \B+d\Lambda$. In the case of exact currents, we have $\B=\extd\mathcal C$ and obviously $\mathcal C$ is not constrained by field equations, thus the corresponding charge is conserved off-shell.
Using the stokes theorem similar to \eqref{softchargememory} we have,
 \begin{align}\label{coexactmemroy}
     Q^{\text{\tiny coexact}}[{i^+}]-Q^{\text{\tiny coexact}}[{i^0}]&=
      \int_{\mathcal{I}^+}\partial_u{\mathcal J}^{ru}=\int_{S^4}\lambda_j \mathcal{D}_i B^{ij}\Big|_{u=-\infty}^{u=+\infty}\\
     Q^{\text{\tiny exact}}[{i^+}]-Q^{\text{\tiny exact}}[{i^0}]&
     =\int_{\mathcal{I}^+}\partial_u\mathcal{K}^{ru}=\int_{S^4}\lambda_u\,{\mathcal D}_i\left({\mathcal D}_i{C}_r\right)\Big|_{u=-\infty}^{u=+\infty}\label{exactmemroy}
 \end{align}
where $B$ and $C$ are the leading in $1/r$ contributions of $\B$ and $\cal C$ while  $\lambda$ is the leading contribution of the gauge parameter $\Lambda$. In the last equality of \eqref{coexactmemroy} we used the field equations $(\extd\star{\cal H})_u=0$. As we will discuss, the coexact charges are relevant to the string memory effect. As we see in \eqref{exactmemroy}, the difference of the exact charges at past and future of null infinity reduces to the case of electromagnetism \eqref{softchargememory} with $\lambda_u$ and ${\mathcal D}_iC_r$ as new gauge parameter and potential.

\section{Closed string memory effect}\label{sec:3}

In this section, we consider a closed string in  flat spacetime located at the radiation zone of a scattering process, which involves 2-form soft radiation\footnote{We note that while the soft charge analysis of \cite{Afshar:2018apx} is made for 2-form theory in six dimensions, our string probe analysis here is valid for strings in target space in any dimensions. This target space may be a six dimensional one associated with a ten dimensional critical superstrings on a four dimensional compact Ricci flat manifold.} and calculate the response of the string  to the 2-form soft radiation. 
The Polyakov action for a closed string, coupled to a background 2-form field $\B_{\mu\nu}$ is:
\begin{align}\label{polyakov action}
    S=-\frac{1}{4\pi\alpha^\p}\int d\tau d\sigma \sqrt{-\gamma}\left[\gamma^{ab}\partial_aX^\mu\partial_b X_\mu+\epsilon^{ab}\partial_aX^\mu\partial_b X^\nu \B_{\mu\nu}\right]\,.
\end{align}
Variations w.r.t. the worldsheet metric $\gamma_{ab}$ and the target space coordinates $X^\mu$ give the following constraint and equations of motion,
\begin{align}
    &(\partial_\pm r)^2-2\ell_s\partial_\pm r+(\partial_\pm X_i)^2=0\,\label{constraint},\\
     &   \partial_+\partial_-X^i=-\ell_s\partial_u{\B}^{ij} X^{\p j}\,.\label{closed eom}
\end{align}
where we used the light-cone gauge $u=\ell_s\tau$ for the coordinate system $(u,r,X^i)$, $i=1,2,\cdots,D-2$ with $u=t-r$, the conformal gauge choice $\gamma_{ab}=e^{\omega}\eta_{ab}$ for the worldsheet metric and the radial gauge $\B_{r\mu}=0$ for the background 2-form field. The $\B$-term is a topological term (it is independent of the worldsheet metric $\gamma^{ab}$) and does not contribute to the  constraint \eqref{constraint} which determines the non-zero-mode part of the $r$-coordinate.

In our analysis, we neglect variation of $\B$ around the string; we assume  $\partial_\sigma\B=0$, that is, the string length is much smaller than variation of $\B$.
The solution to equations of motion \eqref{closed eom} in presence of small $\B$-field  can be related to a solution with vanishing  $\B$-field,
\begin{equation}\label{Y equation}
    \partial_-\partial_+ Y^i=0\,,
\end{equation}
whose general solution is
\begin{equation}\label{gensol}
    Y^i(\tau,\sigma)=Y^i_{L}(\tau-\sigma)+Y^i_{R}(\tau+\sigma)\,,
\end{equation}
where $Y^i_{L,R}(x)=Y^i_{L,R}(x+2\pi)$. The general solution \eqref{gensol} subject to these boundary conditions, takes the following form,
\begin{equation}\label{closed mode exp}
    Y^i(\tau,\sigma)=y_0^i+\alpha^\p p^i\tau+i\sqrt{\frac{\alpha^\p}{2}}\sum_{n\neq 0}\frac{1}{n}\left(\alpha^i_n e^{-in(\tau-\sigma)}+
    \tilde{\alpha}^i_n e^{-in(\tau+\sigma)}
\right)\,.\end{equation}
Next, consider 
\begin{equation}\label{X def Y}
    X^i(\tau,\sigma)\equiv [\delta^{ij}+\frac{1}{2}\B^{ij}(u)]Y^j_L+     [\delta^{ij}-\frac{1}{2}\B^{ij}(u)]Y^j_{R}\,,
\end{equation}
which satisfies
\begin{equation}
    \partial_+\partial_-X^i=-\ell_s \partial_u{\B}^{ij}X'^j+\frac{\ell_s^2}{2}\partial^2_u{\B}^{ij}(Y^j_{L}-Y^j_{R})+\ord{\B^2}\,.
\end{equation}
Assuming that  $\B$ is \emph{soft}, \emph{i.e.} its frequency is much smaller than the string frequency $\ell_s^{-1}$, the second term is negligible and $X^i$ satisfy equations of motion \eqref{closed eom}. Therefore, the closed string solution to \eqref{closed eom} (with periodic boundary conditions) for a slowly varying background $\ell_s\partial_u\B\ll \B$ and neglecting $\ord{\B^2}$ terms is
\begin{align}\label{mode exp closed}
  \hspace*{-0.3cm}  X^i=x_0^i+\alpha^\p p^i\tau+i\sqrt{\frac{\alpha^\p}{2}}\sum_{n\neq 0}\frac{1}{n}\Big[(\delta^{ij}+\frac{1}{2}\B^{ij}(u))\alpha^j_n e^{-in(\tau-\sigma)}+
    (\delta^{ij}-\frac{1}{2}\B^{ij}(u))\tilde{\alpha}^j_n e^{-in(\tau+\sigma)}\Big]\,.
\end{align}
where $\alpha^i_n$ and $\tilde{\alpha}^i_n$ are the oscillator modes of $Y^i_L$ and $Y^i_R$ in \eqref{closed mode exp}, respectively.

\paragraph{Classical closed string memory effect.} The memory effect is the imprint of the background field on the string, found by comparing the state of the string at far future and far past.
Defining
\begin{equation}\label{Theta def}
    \Delta\B_{ij}=\B_{ij}(u\to+\infty)-\B_{ij}(u\to-\infty)\,,
\end{equation}
the closed string memory effect is
\begin{align}\label{classical memory closed}
  \alpha^i_n(u\to+\infty)&=(\delta^{ij}+\frac{1}{2}\Delta\B^{ij})\alpha^j_n(u\to-\infty)\\
       \tilde{\alpha}^i_n(u\to+\infty)&=(\delta^{ij}-\frac{1}{2}\Delta\B^{ij})\tilde{\alpha}^j_n(u\to-\infty)\\
       p^i(u\to+\infty)&= p^i(u\to-\infty)\label{closedmemory}
\end{align}
plus terms of $\ord{\B^2}$ and $\ord{\ell_s\partial_u\B}$ which are negligible.
The center of mass motion is unaltered, while oscillation modes are rotated, with a relative minus sign between left- and right- modes.

\subsection{Quantum closed string memory}
To study the quantum closed string memory, we start from \eqref{classical memory closed}-\eqref{closedmemory} and quantize the system\footnote{Quantum operators are not hatted in this section.}. We employ a Hamiltonian approach, perform the calculations in the light-cone gauge and treat the soft background  as a perturbation. Using proposition \ref{canonic}, we first introduce a canonical transformation that turns the initial and the final Hamiltonian into  similar forms.
The light-cone Hamiltonian for the action \eqref{polyakov action} is
\begin{equation} 
H= \frac{\ell_s^2}{4}\int_0^{2\pi} d\sigma\left[(\check{P}_i-\frac{2}{\ell_s^2}X^{\p j}\B_{ij})(\check{P}_i-\frac{2}{\ell_s^2}X^{\p k}\B_{ik}) +\frac{4}{\ell_s^4}( X^{\p i} X^{\p i})\right]\,.
\end{equation}
where $\check{P}_i=\frac{\partial L}{\partial \dot{X}^i}=\frac{2}{\ell_s}\left(\dot{X}_i+X'^j\B_{ij}\right)$ is the canonical momentum of $X^i$. Consider the canonical transformation on the momenta
\begin{equation}
    P_i=\check{P}_i-\frac{2}{\ell_s^2} X^{\prime j}\B_{ij}\,,
\end{equation}
and leave the coordinates  unaltered. The transformed Hamiltonian $K$ is
\begin{equation}\label{Classical K}
    K= \frac{\ell_s^2}{4}\int_0^{2\pi} d\sigma\left[P^iP^i +\frac{4}{\ell_s^4}  X^{\p i} X^{\p i}+\frac{4}{\ell_s^4}X^i X^{\p j} \dot{\B}_{ij}\right]\,.
\end{equation}
The generating function for the transformation is
\begin{equation}
G=X^i(P_i+\frac{1}{\ell_s^2}X^{\p j}\B_{ij})\,.
\end{equation}

To quantize, as usual, we promote the phase space coordinates to quantum operators and impose the commutation relations
\eq{
[X^i(\sigma),P^j(\sigma^\p)]=i\delta^{ij}\delta(\sigma-\sigma^\p)\,,\qquad [\bar{r},p_{\bar{r}}]=i\,,
}{}
where $\bar{r}=\int d\sigma r(\sigma)$\,.  The free string mode expansion \eqref{closed mode exp} then yields the following brackets
\eq{
[\alpha^i_m,\alpha^j_{n}]=m\delta^{ij}\delta_{m+n},\qquad
[\tilde{\alpha}^i_m,\tilde{\alpha}^j_{n}]=m\delta^{ij}\delta_{m+n}\,,\qquad
[x_0^j,p^j]=i\,.
}{closed-string-brackets}
The Hamiltonian is\footnote{One could perform the canonical transformation \emph{after} quantization, which would be a unitary transformation on Hilbert space and operators.
Consider the following unitary transformation
\eq{
U[\B]=\exp{\left(-\frac{i}{\ell_s^2}\int_0^{2\pi}d\sigma\,
X^iX^{\p j}\B_{ij}
\right)}\,.\nonumber
}{unitary}
with its action on momentum operator
\eq{
P^i\to UP^iU^{\dagger}=P^i+\frac{2}{\ell_s^2}X^{\p j}\B_{ij}-\frac{1}{\ell_s^2 }X^j\partial_\sigma\B_{ij}\nonumber
}{}
with the transformed Hamiltonian 
\begin{equation}\label{KKH}
    K=UHU^\dagger+i\dot{U}U^\dagger.\nonumber
\end{equation}
Assuming  the string length is much smaller than the variation of $\B$, $\partial_\sigma\B\simeq 0$, $K$ 
takes the same form as \eqref{Classical K}, but now as a quantum operator. }
\begin{equation}\label{quantum hamiltonian}
    K=K_0+\sum_{n\neq 0}\frac{i}{2n}(\alpha^i_{-n}\alpha^j_{n}-\tilde{\alpha}^i_{-n}\tilde{\alpha}^j_{n}+2\alpha^i_n\tilde{\alpha}^j_{n})\dot{\B}_{ij}
\end{equation}
where $K_0=\frac{1}{2}\sum_{n}\left(\alpha_n^i\alpha_{-n}^i+\tilde{\alpha}_n^i\tilde{\alpha}_{-n}^i\right)$ is the unperturbed Hamiltonian with  $\alpha_0^i=\tilde{\alpha}_0^i=\sqrt{\alpha^\p/2}\,p^i$. 
In our analysis we will drop the intercept (zero point energy) as we implicitly assume that our $X$ modes are a part of a superstring theory where the zero point energy cancels out by the worldsheet superpartners.

For a slowly-varying small $\B$-field background (\emph{i.e.} the adiabatic evolution, in which the perturbation varies much slower than the energy gap among states), the transition amplitude between states of different energy is vanishing. As a result,  the very last term in \eqref{quantum hamiltonian}, which does not commute with the free Hamiltonian $K_0$  gives no contribution to the evolution of states. (The fact that adiabatic evolution can give rise to transitions only in degenerate states has a simple derivation; see appendix \ref{HOSCharged} for the example of a harmonic oscillator subject to soft electromagnetic radiation.). The first couple of terms are identified as the left and right components of the angular momentum operators\footnote{We are using the notation of \cite{Blumenhagen:2013fgp}, where the angular momentum operator is defined as
\begin{equation}
 J^{\mu\nu}=2\int_0^{2\pi}d\sigma X^{[\mu}P^{\nu]} =\ell^{\mu\nu}+E^{\mu\nu}+\tilde{E}^{\mu\nu}\,,\nonumber
\end{equation}
and $\ell^{\mu\nu}=2x_0^{[\mu}p^{\nu]}$.}
\begin{equation}
    E^{ij}\equiv\sum_{n\neq 0}\frac{i}{n}\alpha^i_n\alpha^j_{-n}\,,\qquad
       \tilde{E}^{ij}\equiv\sum_{n\neq 0}\frac{i}{n}\tilde{\alpha}^i_n\tilde{\alpha}^j_{-n},
\end{equation}
and commute with the Hamiltonian. The Hamiltonian is then easily integrated to give the  time evolution operator
\begin{equation}
    U(u_2,u_1)=\exp\left(-\frac{i}{4}(\B_{ij}(u_2)-\B_{ij}(u_1)(E^{ij}-\tilde{E}^{ij})\right)\exp\left(-iK_0(u_2-u_1)\right)\,.
\end{equation}
The $S$ operator which maps $in$ and $out$ states is thus identified as
\begin{equation}\label{S-matrix-closed-string}
    S=\exp\left(-\frac{i}{4}\Delta\B_{ij}(E^{ij}-\tilde{E}^{ij})\right)\,,
\end{equation}
and $\Delta\B_{ij}$ is defined in \eqref{Theta def}. The 2-form memory effect on a closed string is then given by the Heisenberg-picture  evolution of the operators,
\begin{equation}\label{quantum closed memory}
\begin{split}
S^{-1}\alpha^k_m\,S&=+\frac{1}{2}\Delta\B_{ik}\alpha_m^k +\ord{\B^2}\,,\cr
S^{-1}\tilde{\alpha}^k_m\,S&=-\frac{1}{2}\Delta\B_{ik}\tilde{\alpha}_m^k +\ord{\B^2}\,,
\end{split}
\end{equation}
consistent with the classical result \eqref{classical memory closed}.

Eqs.\eqref{quantum closed memory} are our main result of this section. To explore it further, consider the general massless state in bosonic string theory
\begin{equation}
    \zeta_{ij}\alpha^{i}_{-1}\tilde{\alpha}_{-1}^j|0;k\rangle\,.
\end{equation}
The memory effect on this state is the transition in its polarization tensor according to \eqref{quantum closed memory}:
\begin{equation}
    \zeta_{ij}\to \left(\delta^{im}+\frac{1}{2}\Delta\B^{im}\right) \zeta_{mn}\left(\delta^{jn}-\frac{1}{2}\Delta\B^{jn}\right)\,,
\end{equation}
where as usual the dilaton, graviton and $b$-field states are respectively associated with trace, symmetric-traceless and antisymmetric parts of polarization tensor $\zeta_{ij}$:
\begin{equation}
\sqrt{D-2}\varphi\equiv \delta^{ij}\zeta_{ij}\,,\qquad
   b_{ij}=\frac{1}{2}(\zeta_{ij}-\zeta_{ji})\,,\qquad
    h_{ij}=\frac{1}{2}(\zeta_{ij}+\zeta_{ji})-\frac{1}{\sqrt{D-2}}\delta_{ij}\varphi\,.
\end{equation}
Variation of different components are
\begin{subequations}\label{string-theory-conversion-amplutides}
\begin{align}
    \Delta\varphi&=-\frac{1}{\sqrt{D-2}}\Delta\B^{mn}b_{mn}\,,\\
    \Delta b_{ij}&=\frac{\varphi}{\sqrt{D-2}}\Delta\B_{ij}-\Delta\B_{n[i}h_{j]n}\,,\\
    \Delta h_{ij}&=\frac{1}{D-2}\delta_{ij}\Delta\B^{mn}b_{mn}+\Delta\B_{n(i}b_{j)n}\,.
\end{align}
\end{subequations}
Variations of $\varphi$ and $h_{ij}$ depend exclusively on the initial 2-form state $b_{ij}$. On the other hand, variation of the 2-form field comes from  other components, \emph{i.e.} $\varphi$ and $h_{ij}$. This is of course quite expected from 
group theory viewpoint; the contracted product of an antisymmetric tensor and symmetric one is an antisymmetric tensor and the contracted product of two antisymmetric tensors is a symmetric one. That is, the passage of a soft $2$-form converts the massless $b$ state to dilaton or graviton and vice versa, without changing its momentum or energy.


\subsection{Effective field theory analysis}
In this subsection, we will provide a field theoretic explanation of the closed string memory effect. We write the effective field theory for the graviton $h_{\mu\nu}$,   the dilaton $\varphi$ and the Kalb-Ramond 2-form field $b_{\mu\nu}$, on a slowly varying 2-form background. We calculate the transition amplitudes of previous sections by reading the vertices in Feynman rules.

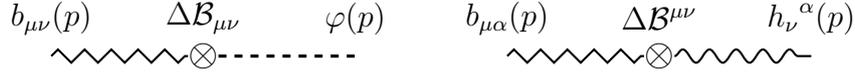
\begin{figure}
    \centering
\begin{tikzpicture}
\usetikzlibrary{snakes}
\draw[snake=zigzag,thick](0,0)--(1.8,0);
\node [thick] at (2,0){\large$\otimes$};
\draw[very thick,dashed] (2.2,0)--(4,0);
\node [] at (0,.5) {$b_{\mu\nu}(p)$};
\node [] at (4,.5) {$\varphi(p)$};
\node [] at (2,.5) {$\Delta\B_{\mu\nu}$};

\begin{scope}[shift={(6,0)}]
\draw[snake=zigzag,thick](0,0)--(1.8,0);
\node [thick,opacity=1] at (2,0){\large$\otimes$};
\draw[thick,snake=snake] (2.2,0)--(4,0);
\node [] at (0,.5) {$b_{\mu\alpha}(p)$};
\node [] at (4,.5) {${h_{\nu}}^\alpha(p)$};
\node [] at (2,.5) {$\Delta\B^{\mu\nu}$};

\end{scope}
\end{tikzpicture}
    \caption{Kinetic mixing vertices. }
    \label{Fig-2}
\end{figure}

The effective action of NSNS sector of string theory in string frame is \cite{Johnson:2003gi}
\begin{align}
    S=\frac{1}{2\kappa^2}\int d^Dx\sqrt{-G}e^{-2\Phi}&\left[R+4\nabla_\mu\Phi\nabla^\mu\Phi-\frac{1}{12}\Hh_{\mu\nu\laa}\Hh^{\mu\nu\laa}
    \right]
\end{align}
where $\kappa^2=8\pi G_N$. For each field we consider a perturbation on top of a background field
\begin{equation}
    G_{\mu\nu}=\bar{G}_{\mu\nu}+h_{\mu\nu}\,,\qquad \B_{\mu\nu}=\bar{\B}_{\mu\nu}+b_{\mu\nu}\,,\qquad \Phi=\bar{\Phi}+\varphi
\end{equation}
We assume that background fields satisfy their own classical equations of motion, thus, only terms of second and higher order in perturbation fields appear in the action. The second-order terms are kinetic terms. We assume that $\bar{G}_{\mu\nu}=\eta_{\mu\nu}$ and $\bar{\Phi}=0$
and neglect second order terms in background $\B$-field. The second order action in Einstein frame
\footnote{
The Einstein frame action is
\begin{align}
    S=\frac{1}{2\kappa^2}\int d^Dx\sqrt{-G}&\left[R-\frac{4}{D-2}\nabla_\mu\Phi\nabla^\mu\Phi-\frac{1}{12}e^{-8\Phi/(D-2)}\Hh_{\mu\nu\laa}\Hh^{\mu\nu\laa}
    \right]\,.\nonumber
\end{align}
} becomes
\begin{equation}
    S^{(2)}=\frac{1}{2\kappa^2}\int d^Dx\left[\mathcal{L}^{(2)}_h+\mathcal{L}^{(2)}_b+\mathcal{L}^{(2)}_\varphi\right]
\end{equation}
where
\begin{subequations}
\begin{align}    \mathcal{L}^{(2)}_h&=-\frac{1}{2}\nabla_\mu h_{\nu\rho}\nabla^\mu h^{\nu\rho}+\nabla_\mu h_{\nu\rho}\nabla^\nu h^{\mu\rho}-\nabla_\mu h\nabla_\rho h^{\mu\rho}+\frac{1}{2}\nabla_\mu h\nabla^\mu h\\
    \mathcal{L}^{(2)}_b&=-\frac{1}{12}(\di b)_{\mu\nu\laa}(\di b)^{\mu\nu\laa}\\
    \mathcal{L}^{(2)}_\varphi&=-\frac{4}{D-2}\nabla_\mu\varphi\nabla^\mu\varphi
\end{align}
\end{subequations}
There are also a couple of kinetic mixing terms (interaction with background $\B$-field)  at second order due to the background $\Hh$-field, as depicted in Figure \ref{Fig-2}
\begin{align}
    \mathcal{L}^{(2)}_{b\varphi}&=\frac{4}{3(D-2)}\varphi\bar{\Hh}_{\mu\nu\laa}(\di b)^{\mu\nu\laa}\\
        \mathcal{L}^{(2)}_{bh}&=-\frac{1}{6}\left(3h^{\alpha\beta}\bar{\Hh}_{\alpha\nu\laa}{(\di b)_{\beta}}^{\nu\laa}+\frac{1}{2}h\bar{\Hh}_{\mu\nu\laa}(\di b)^{\mu\nu\laa}\right)
\end{align}

\paragraph{Gauge fixed action; TT gauge.}
We impose transversality condition both on graviton and 2-form field:
\begin{equation}\label{transverse}
   \nabla^\mu h_{\mu\nu}=0\,,\qquad \nabla^\mu b_{\mu\nu}=0\,.
\end{equation}
In addition, the temporal components can be set to zero by using the residual gauge symmetry while the metric perturbation is set to traceless:
\begin{equation}\label{traceless}
   \eta^{\mu\nu}h_{\mu\nu}=0\,,\qquad h_{u\alpha}=0\,, \qquad b_{u\alpha}=0\,.
\end{equation}
Equations \eqref{transverse} and \eqref{traceless} comprise $2D$ conditions on graviton  and $2D-3$ conditions on the 2-form field to give the right degrees of freedom.

We implement canonical quantization of the fields on constant-$u$ surfaces. Although $u$ is a timelike coordinate, constant-$u$ surfaces, where outgoing wavefronts lie, are null hyperplanes. The canonical momenta are the following
\begin{align}
    \Pi_{(\varphi)}(\x)&=\frac{\partial\mathcal{L}}{\partial\dot{\varphi}}=\frac{4}{\kappa^2(D-2)}\partial_r\varphi\\
        \Pi^{ab}_{(b)}(\x)&=\frac{\partial\mathcal{L}}{\partial\dot{b}_{ab}}=-\frac{1}{2\kappa^2}H^{uab}\\
        \Pi^{ab}_{(h)}(\x)&=\frac{\partial\mathcal{L}}{\partial\dot{h}_{ab}}=\frac{1}{2\kappa^2}\nabla_rh^{ab}
\end{align}
The interaction-picture free fields are\footnote{The completeness relations are
\begin{align}
    \sum_{s}\zeta^s_{ab}(\pii)\zeta^{s\ast}_{cd}(\pii)&=\delta_{a[c}\delta_{d]b}-2\pii_{[a}\delta_{b][d}\pii_{c]}/|\pii|^2\,,\nonumber\\
    \sum_{s}\xi_{ab}(\pii,s)\xi^{s\ast}_{cd}(\pii,s)&=\delta_{a(c}\delta_{d)b}-2\pii_{(a}\delta_{b)(d}\pii_{c)}/|\pii|^2\,.\nonumber
\end{align}
}
\begin{align}
   \varphi(\x,t)&=\frac{\kappa}{2}\sqrt{D-2}\int \di^{D-1}p\frac{1}{\sqrt{2p_r}}\left(\boldsymbol{\varphi}_{\pii}e^{ip\cdot x}+\boldsymbol{\varphi}^\dagger_{\pii}e^{-ip\cdot x}\right)\\
      b_{ab}(\x,t)&=\kappa\sum_s\int \di^{D-1}p\frac{1}{\sqrt{2p_r}}\left(\zeta_{ab}(\pii,s)\,\mathbf{b}^{s}_{\pii}\,e^{ip\cdot x}+\zeta^{\ast}_{ab}(\pii,s)\,\mathbf{b}^{s\dagger}_{\pii}\,e^{-ip\cdot x}\right)\,,\\
          h_{ab}(\x,t)&=\kappa\sum_s\int \di^{D-1}p\frac{1}{\sqrt{2p_r}}\left(\xi_{ab}(\pii,s)\,\mathbf{h}^{s}_{\pii}\,e^{ip\cdot x}+\xi^{\ast}_{ab}(\pii,s)\,\mathbf{h}^{s\dagger}_{\pii}\,e^{-ip\cdot x}\right)\,.
\end{align}
The quantum operators satisfy
\begin{subequations}
\begin{align}
    [\boldsymbol{\varphi}_\pii,\boldsymbol{\varphi}^\dagger_{\textbf{q}}]&=\delta^{D-1}(\pii-\textbf{q})\,\\
        [\mathbf{b}^r_\pii,\mathbf{b}^{s\dagger}_{\textbf{q}}]&=\delta^{rs}\,\delta^{D-1}(\pii-\textbf{q})\,,\qquad s=1,\cdots(D-2)(D-3)/2\\
    [\mathbf{h}^r_\pii,\mathbf{h}^{s\dagger}_{\textbf{q}}]&=\delta^{rs}\,\delta^{D-1}(\pii-\textbf{q})\,,\qquad s=1,\cdots,D(D-3)/2\,.
\end{align}
\end{subequations}
The free Hamiltonian is
\begin{equation}
    H=\int d^{D-1}p\,\omega_\pii\left(\boldsymbol{\varphi}^\dagger_\pii \boldsymbol{\varphi}_\pii+\sum_s \mathbf{b}^{s\dagger}_\pii \mathbf{b}^{s}_\pii+\sum_s\mathbf{h}^{s\dagger}_\pii \mathbf{h}^{s}_\pii\right)\,,\qquad \omega_\pii=p_u=\frac{p_r^2+p_i^2}{2p_r}\,.
\end{equation}

\paragraph{Two-form--dilaton conversion amplitude.}

The time evolution operator (interaction Hamiltonian) at first order in $\B$ is
\begin{equation}
    -i\int du H_I(u)=-\frac{2i}{3(D-2)\kappa^2}\int d^D x\varphi(x)\bar{\Hh}_{\mu\nu\laa}(x)(\di b)^{\mu\nu\laa}(x)\,.
\end{equation}
Fourier transform on spatial dimensions and using free field expansions gives
\begin{equation}\label{b phi vertex}
\frac{3}{\sqrt{D-2}}\int du \bar{\Hh}_{uij}(u)\sum_{s}\int \frac{d^{D-1}p}{2p_r}\left(p^{[u}\zeta^{ij]\ast}(\pii,s)\boldsymbol{\varphi}_{\pii}\mathbf{b}^{s\dagger}_\pii\,-p^{[u}\zeta^{ij]}(\pii,s)\boldsymbol{\varphi}^\dagger_{\pii}\mathbf{b}^{s}_\pii\,\right)\,.
\end{equation}
The first integral picks out the soft mode of the background field
\begin{equation}
    \int du\bar{\Hh}_{uij}=\tilde{\Hh}_{uij}(\omega\to 0)=\Delta\B_{ij}\,.
\end{equation}
The transition amplitude is hence proportional to the intensity of the soft 2-photon. 

The simplest example the matrix element $\mathcal{M}(b_{\mu\nu}\to\varphi)$, between an in-going\footnote{Outgoing radial particles are beyond our scope. They move on constant $u$ world-lines, where ``time stops''. In other words, the out-going field is only a function of $u=t-r$, whereas the in-going field is only a function of $v=t+r=u+2r$. So, for the out-going particle $p_r=0$ and the conjugate momentum vanishes.} 2-form particle with momentum $p _\mu=(p,2p,0,\cdots,0)$ and polarization tensor $\zeta_{ij}(\pii,s)$, and a scalar particle of the same momentum. The amplitude according to \eqref{b phi vertex} is
\begin{equation}
 \mathcal{M}(b_{\mu\nu}\to\varphi)=\frac{-1}{\sqrt{D-2}}\,\Delta \B_{ij}\zeta^{ ij}(\pii,s)\,.
\end{equation}

\paragraph{Two-form--graviton conversion amplitude.}
The time evolution operator (interaction Hamiltonian) is
\begin{align}\label{b h vertex}
-i\int du H_I(u)=\frac{i}{4\kappa^2}\int d^Dx h^{\alpha\beta}(x)\bar{\Hh}_{\alpha\nu\laa}(x){(\di b)_{\beta}}^{\nu\laa}(x)\,,
\end{align}
and in terms of quantum operators it is
\begin{align}
   \frac{3}{4} \int du &\bar{\Hh}_{uij}(u)\times\nn\\
\hspace*{-8mm}&\sum_{s,w}\int \frac{d^{D-1}p}{2p_r}\left({\xi_\beta}^{[u}(\pii,w)p^{i}{\zeta^{j]\beta\ast}}(\pii,s)\, \mathbf{h}^w_{\pii}\mathbf{b}^{s\dagger}_\pii\,-{\xi_\beta}^{\ast[u}(\pii,w)p^{i}{\zeta^{j]\beta}}(\pii,s)\,\mathbf{h}^{w\dagger}_{\pii}\mathbf{b}^{s}_\pii\,\right).
\end{align}
The amplitude for incoming graviton of momentum $p$, to convert to two-form $b$ is
\begin{equation}
 \mathcal{M}(h_{\mu\nu}\to b_{\mu\nu})=-\Delta \B_{ij}\xi^{m[i}(\pii,s)\zeta^{ j]m\ast}(\pii,w).
\end{equation}
The above of course matches with our string theory computation and result \eqref{string-theory-conversion-amplutides}.

\section{Open string memory effect}\label{sec:4}
In the previous section we showed how the closed string memory effect was a consequence of invariance of the worldsheet or the corresponding low energy effective action, under $\Lambda$-gauge transformations, $\B\to \B+d\Lambda$. For the open string case and due to the presence of worldsheet endpoints the $\Lambda$ gauge symmetry should be modified by the addition of the boundary  1-form gauge field \cite{Witten:1995im, SheikhJabbari:1999ba}.	The action \eqref{polyakov action} augmented by the boundary term
\begin{equation}
S_b=\frac{2}{\ell_s^2}\int d\tau\dot{X}^I\A_I\,,
\end{equation}
where $I,J=1,\cdots,p$ denote the spatial directions 
along the D$_p$-brane and $\A_I$ is the U(1) gauge field on it, leads to the total open string world-sheet action.   Variation of this action, and then fixing the light-cone gauge yields equations of motion \eqref{closed eom} and a boundary term,
\begin{align}
-\frac{2}{\ell_s^2}\int d\tau &\left(\gamma^{\sigma b}(\delta X^I\partial_b X_I)+\epsilon^{\sigma b}\delta X^I(\partial_b X^J \B_{IJ}+\B_{Iu})+\delta X^I\frac{d}{d\tau}\A_I\right)\,,\quad b=\tau,\sigma\,.
\end{align}
which results in the mixed (Neumann and Dirichlet) boundary conditions for directions along the brane \cite{SheikhJabbari:1997yi, SheikhJabbari:1999xd}
\begin{equation}\label{Neumann b.c.}
X^\p_I+\F_{IJ}\dot{X}^J+\ell_s\F_{Iu}=0\,.
\end{equation}

The boundary action introduces a coupling of the boundary $\partial\Sigma$ of the open string to the gauge field on the D-brane which is invariant under the new $\Lambda$-gauge transformations,
\be
\delta\B_{IJ}=2\partial_{[I}\Laa_{J]},\qquad \delta \A_I=\Laa_I
\ee  
In other words, the combination
\be\label{calF}
{\cal F}\equiv\B-d\A
\ee
is invariant under  $\Lambda$-gauge transformation as well as under $\lambda$-gauge transformation, $\A\to \A+d\lambda$.

As in the closed string case,  in the light-cone gauge $u=\ell_s\tau$ by fixing the radial gauge $\B_{r\mu}=0$, we find
\begin{align}\label{mode exp general}
    X^I&=x_0^I+2\alpha^\p p^I\tau+w^I \sigma\nn\\&+i\sqrt{\frac{\alpha^\p}{2}}\sum_{n\neq 0}\frac{1}{n}\left(\Big[\delta^{IJ}+\frac{\B^{IJ}(u)}{2}\Big]\alpha^J_n e^{-in(\tau-\sigma)}+
    \Big[\delta^{IJ}-\frac{\B^{IJ}(u)}{2}\Big]\tilde{\alpha}^J_n e^{-in(\tau+\sigma)}\right)
\end{align}
Next, we impose the boundary condition \eqref{Neumann b.c.}:\footnote{Here we considered the case with both ends of the open string with the same boundary conditions. In principle one can consider cases where the two ends have different Neumann, Dirichlet or mixed boundary conditions, e.g. as in \cite{SheikhJabbari:1997yi}.}
\begin{equation}\label{mode neumann}
    (1-\F)^{IJ}\left(1+\frac{\B}{2}\right)_{JL}\alpha^L_n=
    (1+\F)^{IJ}\left(1-\frac{\B}{2}\right)_{JL}\tilde{\alpha}^L_n\,,
\end{equation}
Since $\B_{IJ}$ and $\F_{IJ}$ have time dependence, \eqref{mode neumann} is in general not consistent with constant $\alpha^i_n$ and $\tilde{\alpha}^i_n$ (which is required by equations of motion). In a slowly varying and small 2-form $\B_{IJ}$ background, this can, however, be remedied if
 \begin{equation}\label{F-B-Lambda}
     \F^{IJ}(u)-\frac{1}{2}\B^{IJ}(u)=C^{IJ}\,,
 \end{equation}
where $C$ is a constant 2-form. We can further choose  $C=0$ by making an appropriate $\Lambda$-gauge transformation on $\B$ such that, the boundary condition \eqref{Neumann b.c.} yields
\begin{equation}\label{alpha-tilde-alpha-open}
     \alpha_n^I=\tilde{\alpha}_n^I\,.
\ee
Eq.\eqref{Neumann b.c.} also fixes the zero mode part, and the mode expansion becomes
\begin{align}\label{open-mode-expansion}
\hspace*{-4mm}X^I(\tau,\sigma)=x_0^I+2\alpha^\p p^I\tau-2\alpha^\p\F^{IJ}p^J\sigma&-2\alpha^\p\ell_s \F_{Iu}\sigma\nn\\
    &+\sqrt{2\alpha^\p}\sum_{n\neq 0}\frac{e^{-in\tau}}{n}\left(i\cos n\sigma\delta^{IJ}-\sin n\sigma\F^{IJ}\right)\alpha^J_n\,,
\end{align}
where the $\ord{\B^2}$ and higher order terms are neglected. 

The analysis above and in particular \eqref{alpha-tilde-alpha-open} and \eqref{open-mode-expansion} shows that the effects of adiabatically time-varying $\B$-field on open string is completely different than on closed string. Specifically, the mode expansion coefficients $\alpha_n^i$ and the Hamiltonian of the system in the appropriate canonical frame are both time independent. 
This is essentially the same as the open string mode expansion in a constant $\B$-field background \cite{SheikhJabbari:1997yi, SheikhJabbari:1999vm, Ardalan:1999av, SheikhJabbari:1999ba,Chu:1998qz,Chu:1999gi}. The effects of the time-variation of the $\B$-field may be seen in the center of mass motion of the string 
\be
{\bar X}^I(u)=\frac{1}{\pi}\int_0^{\pi} d\sigma\ X^I(\tau,\sigma)=x_0^I+2\alpha^\p p^I\tau-\pi\alpha^\p\F^{IJ}p^J-\pi\alpha^\p\ell_s \F_{Iu},
\ee
in the last two terms. We will discuss this further in the following subsection.

\subsection{Quantum treatment}

Having the mode expansion we can readily quantize the open string by imposing \cite{Ardalan:1999av, SheikhJabbari:1999vm, Chu:1998qz,Chu:1999gi, Seiberg:1999vs}
\be
[x^I_0, x^J_0]=i\pi\alpha'\F^{IJ},\qquad [x^I_0, p^J]=i\delta^{IJ},\qquad [\alpha^I_n, \alpha^J_m]=n\delta^{IJ} \delta_{m+n,0}.
\ee
As we see the effects of the adiabatically changing background $\B$-field has appeared only in the noncommutativity of the $x_0^i$ coordinates. Note that $x_0^i$ are basically the coordinates of the D$_p$-brane the open string endpoint is attached to.  Since the Hamiltonian is not affected by the background $\B$-field in the open string case we do not have a memory effect in the usual sense. Nonetheless, open strings in the adiabatically changing $\B$-field background behaves as an electric dipole \cite{SheikhJabbari:1999vm}, whose dipole moment is changing in time:
\begin{equation}
\Delta d^I\equiv  d^I (u\to+\infty)- d^I (u\to-\infty)=2\pi\alpha^\p\Delta\B^{IJ}p^J,
\end{equation}
where $d_I(u)=\langle X^I(\sigma=\pi, u)-X^I(\sigma=0, u)\rangle$. This result can be also understood from the effective field theory of the open strings, which is the Born-Infled theory residing on the brane. In presence of an adiabatically changing $\B$-field we are dealing with a noncommutative gauge theory \cite{SheikhJabbari:1998ac,Douglas:1997fm} with a slowly varying noncommutativity parameter (see also \cite{Bachas:2002jg} for a related analysis). As it is known in the noncommutative field theories the kinetic term is not affected by noncommutativity, e.g. see \cite{Micu:2000xj} and the effects of noncommutativity appear only in interaction terms which is not captured in the usual memory effect described in previous sections. 

\subsection{D-brane probes and boundary states}
D-barnes are part of the spectrum of the string theory. They appear by requiring T-duality in theories of open strings \cite{Polchinski:1996fm,Polchinski:1996na} by the fact that under T-duality, Neumann and Dirichlet boundary conditions are interchanged.
Besides this open string description \cite{Polchinski:1995mt}, it is known that D-branes can be represented as a coherent (bound) state of closed strings they can emit. The amplitude of the closed string emission is given through the boundary state \cite{DiVecchia:1997vef},  for a review of constructing these states  see \cite{DiVecchia:1999mal,DiVecchia:1999fje} and references therein. 

The open string satisfies the boundary condition \eqref{Neumann b.c.} at its end along the D-brane world-volume and along the transverse directions we have $X^i(\sigma=0)=x^i_0$ for $i=p+1,\cdots,d-1$.
Boundary state for D$_p$-branes in a constant $\B$-field background has also been worked out and studied \cite{DiVecchia:1999uf,Arfaei:1999jt}. One may generalize the analysis of \cite{Arfaei:1999jt} to adiabatically changing $\B$-field background, which should satisfy\footnote{Here we prefer covariance and do not to go to the light-cone gauge.}
\be
\left(\partial_\tau X_\mu+\F_{\mu\nu} \partial_\sigma X^\nu\right)_{\tau=\tau_0}|\mathscr{B}(\tau_0)\rangle=0\,.
\ee
We note that the boundary state $|\mathscr{B}(\tau_0)\rangle$ should satisfy the above condition at the given arbitrary time $\tau_0$, which in the light-cone gauge $\tau_0=u_0/\ell_s$. That is, $|\mathscr{B}(\tau_0)\rangle$ is giving the amplitude of closed string mode emissions from a D-brane at time $\tau_0$. In a static background like the cases analyzed in \cite{DiVecchia:1999uf, Arfaei:1999jt}, the $\tau_0$ dependence of the boundary state  appears simply through $e^{in\tau_0}$ dependence of the string oscillator modes,  while in our case there is an extra dependence due to the slowly varying background $\B$ field.
Since the analysis is  similar to the static case where $\F$ is constant, we skip the details of computation and quote the final result:
\begin{equation}\label{boundary-state}
    |\mathscr{B}(\tau_0)\rangle=\exp\left(-\sum_{n=1}\frac{e^{2in\tau_0}}{n}\alpha^{\mu}_{-n}D_{\mu\nu}(\tau_0)\tilde{\alpha}^{\nu}_{-n}\right)|0\rangle\,,\qquad D_{\mu\nu}=(\mathcal{Q}_{\alpha\beta}(\tau_0),-\delta_{AB})\,.
\end{equation}
where $\alpha,\beta$ label all directions parallel to the brane, while those normal to the brane are labelled by $A,B$. The matrix $\mathcal{Q}(\tau_0)$ is defined as
\begin{equation}
    \mathcal{Q}(\tau_0)=(1-\F(\tau_0)+\B(\tau_0)/2)^{-1}(1+\F(\tau_0)-\B(\tau_0)/2)=1+2\F(\tau_0)-\B(\tau_0)+\cdots\,,
    \end{equation}
where the braces denote higher orders in $\B$ or $\F$. Having the D$_p$-brane boundary state we can use it as the probe to explore the memory effect associated with passage of a $\B$-field. As the first example, let us compute the action of the closed string memory $S$-matrix \eqref{S-matrix-closed-string} on this state:
\begin{align}
    S|\mathscr{B}\rangle=\left\{1+\frac{1}{2}\sum_{n=1}^{\infty}\frac{e^{2in\tau_0}}{n}\left(\alpha^\alpha_{-n}\left(\Delta \B\right)_{\alpha\beta}\tilde{\alpha}^\beta_{-n}-\alpha^\alpha_{-n}\left(\Delta \B\right)_{\beta\alpha}\tilde{\alpha}^\beta_{-n}-\left({2\Delta \B}\right)_{\alpha\beta}\alpha^\alpha_{-n}\tilde{\alpha}^\beta_{-n}\right)+\cdots\right\}|\mathscr{B}\rangle,
\end{align}
where braces denote terms second or higher order in $\B, \F$. We therefore have
\eq{
S|\mathscr{B}\rangle=|\mathscr{B}\rangle+\ord{\B^2}\,.
}{}
The boundary state is unchanged at first order in background fields. This result may be understood as follows. 
The D$_p$-brane in a slowly-varying $\B$-field background is a non-marginal bound state of D$_p$ and lower dimensional D$_{(p-2n)}$-branes where $n$ rank of the $\B$ field along the brane, much like the constant $\B$-field case \cite{SheikhJabbari:1997yi}. The  mass density of the brane is $\frac{1}{g_s}\sqrt{\det{(1+\F)}}\simeq \frac{1}{g_s}(1+{\cal O}(\F^2))$ where $g_s$ is the string coupling. This is compatible with the softness of the passing $B$-field wave. The ($p-2n$)-form RR charge density of carried by the bound state  is proportional to $\F^n$. Therefore, in our approximation only  $n=1$ case is remaining. The change in this D$_{(p-2)}$-brane charge density is then proportional to $\Delta\B$. We note, however, that to see this RR charge density from the boundary state one needs to go beyond the bosonic sector discussed above, and to consider the superstring case and include fermionic degrees of freedom. Moreover, using the boundary state \eqref{boundary-state} one can study scattering of two such D$_p$-branes off each other, where the $|in\rangle$ and $|out\rangle$ boundary states differ in their value of the $\F$ field.

\section{Discussion}\label{sec:5}

In this work, we continued our analysis of the $p$-form soft charges focusing on the $2$-form case and studied how these charges can be probed. In particular, we considered the natural probes of a $2$-form, the strings, and studied the 2-form memory imprinted on them due to the passage of a 2-form wave-packet. The string memory effect we analyzed has some particular features not shared by previously studied memory effects: the memory is encoded in an internal excitation mode of the probe, here the strings, and that the probe itself can be a massless object.\footnote{While in this work we focused on the internal string memory effect, we note that the center of mass of massive string modes can exhibit the same external memory effects as usual particle probes under gravitational waves can have, e.g. the displacement \cite{Bachas:2002jg} or super-boost memories.}  We discussed string memory effect for closed and open strings and mentioned that the latter can also be analyzed through D-brane probes, using boundary state formulation. In our analysis here, we demonstrated how the basic idea works for bosonic strings and of course we expect the same analysis to be worked through, almost verbatim, for superstring case, either in RNS or GS formulations \cite{Polchinski:1998rr}.

While the $p$-form soft charge analysis of \cite{Afshar:2018apx} was carried out for $2p+4$ dimensional theories, our analysis here was made for strings in generic dimension $D$. The six dimensional strings (for $p=2$ case) may then be viewed as a critical 10 dimensional superstring compactified on a 4d manifold. In this case, the strings may have winding modes and the $\B$-field may have legs along the compact direction. It is interesting to check how these may affect/appear in the string memory.

As discussed in \cite{Afshar:2018apx}, there are two classes of $p+1$-form charges for $p\geq 1$, the coexact charges and the exact soft charges. The former  is a direct generalization of the $p=0$ electromagnetic soft charges case and as we discussed here these are the charges relevant for the closed string memory effect. To discuss relevance of exact soft charges to the memory effect, we recall that the exact configurations $\B=\extd\mathcal C$ have vanishing field strength but can in principle be probed by open strings where the worldsheet has a boundary. In the case that open strings end on D-branes, \emph{cf.} section \ref{sec:4}, the brane may be viewed as the ``boundary of the spacetime'' and the ${\cal C}$ field which may be viewed as the edge-mode associated to the two-form, can be identified with the gauge field on the brane. The corresponding memory effect is then expected to essentially reduce to the electromagnetic memory effect on the brane.

Our discussions here can be readily extended to generic $p$-form memory using $(p-1)$-brane probes. Again we expect to have a similar memory effect sharing the features of string memory effect. This $p$-brane memory effect can then be of relevance for black hole information problem within the membrane paradigm viewpoint \cite{Grumiller:2018scv}. 

Here and in \cite{Afshar:2018apx} we studied two corners of the $p$-form ``IR-triangle'' \cite{Strominger:2017zoo}. It is desirable to complete this by studying (1) more direct relation of $p$-form soft charges of \cite{Afshar:2018apx} and the memory effect and, (2) study the $p$-form soft theorems and connecting it to the other two corners.  For some relevant analysis especially within string theory framework see \cite{DiVecchia:2015oba,DiVecchia:2015srk,DiVecchia:2017gfi,Sen:2017xjn,Laddha:2017ygw}. As general comment on soft theorems we note that for any observable $X$  on the celestial sphere, the difference in its early and late retarded time values $\Delta X$ is given by
\eq{
\Delta X=\int_{u=-\infty}^{u=+\infty}\extd u \,\dot X=\int_{u=-\infty}^{u=+\infty}\extd u\int_{\omega=-\infty}^{\omega=+\infty} i\omega\extd\omega \,e^{i\omega u} \,\tilde{X}(\omega)=\lim_{\omega\to0}i\omega\, \tilde{X}(\omega)\,.}{}
This shows that requiring a nontrivial $\Delta X$ implies a nontrivial simple pole in $\tilde X$.

\acknowledgments

We would like to thank Paolo Di Vecchia, Vahid Hosseinzadeh, Reza Javadinezhad, Massimo Porrati, Hesam Soltanpanahi and Alexander Zhiboedov for fruitful discussions.
H.A. and M.M.Sh-J. are partially supported by the junior research chair in black hole physics of Iranian NSF, project no 951024. We  also acknowledge the ICTP NT-04 network scheme. H.A. would like to gratefully acknowledge the support of the CERN Theory Department and the Iran-Austria IMPULSE project grant during his stay at CERN and TU Wien while this work was being prepared.

\appendix

\section{Absence of electromagnetic memory for harmonic oscillator probes}\label{HOSCharged}

As we reviewed in section \ref{sec:2}, adiabatically changing background electric field is encoded as the change in momentum of a charged free particle probing the background, the electromagnetic kick memory. The kick on the particle is accumulated over the particle trajectory from far in the past to far in future. One may wonder whether this momentum shift accumulation also happens for a probe which has perioidic trajectories, like a harmonic oscillator. As we show below the answer is negative.
\paragraph{Classical treatment.}
 The classical Hamiltonian of a charged harmonic oscillator in the radial gauge $A_r=0$ is,
\begin{align}\label{HOS}
    H=H_0+\frac{1}{2}m\omega^2(X_iX^i+r^2)\,,
\end{align}
where $H_0$ is the Hamiltonian for the free charged particle \eqref{classical hamilton}. We now want to show that the zero-frequency mode of electromagnetic radiation leaves no memory on a charged harmonic oscillator.
By defining $\bar{p}_i\equiv p_i/\sqrt{m\omega}$, $\bar{x}^i\equiv\sqrt{m\omega}x^i$ and  $\bar{\A}_i\equiv q\,\A_i/\sqrt{m\omega}$, the Hamiltonian \eqref{HOS} takes a simple form,
\eq{
K=\frac{\omega}{2}\left(\bar{p}_i^2+\bar{x}_i^2\right)-\bar{x}^i\dot{\bar{\A}}_i\,.
}{}
Equations of motion for transverse momenta are
\eq{
\ddot{\bar{p}}_i+\omega^2\bar{p}_i=q\,\dot{\bar{\A}}_i
}{2nd ord}
The homogeneous solution is
$
\bar{p}_i(u)=\beta(u)+c.c.$ where $\beta(u)=\hat{\beta} e^{i\omega u}
$ 
and $\hat{\beta}$ is a fixed complex number. Suppose that we are interested in the solution in interval $u\in[-L,L]$. Divide the interval into $N$ segments and define $u_n=L(2n/N-1)$. Evolution of the  homogeneous solution is given by the phase shift  $\beta(u_{n+1})=U_N \beta(u_n)$, where $U_N=\exp(2iL\omega /N)$.

Suppose $\A_i$ has a small shift at each step, such that $\dot{\bar{\A}}_i=\epsilon_i\delta(u-u_n)/N$, where $\epsilon_i=q\,(\A_i^+-\A_i^-)/\sqrt{m\omega}$. Since the momentum shift in each interval is 
\eq{
\lim_{u\to u_n^+}\beta_i(u)-\lim_{u\to u_n^-}\beta_i(u)=-i\epsilon_i/N,
}{}
we arrive at the following recursive formula 
\begin{equation}
\beta(u_{n+1})=U_N\beta(u_n)-i\epsilon_i/N\,,\qquad n=0,\cdots,N,
\end{equation}
and hence
\begin{equation}
\beta(L)=(U_N)^N\,\beta(-L)-\frac{i}{N}\epsilon_i\sum_n^{N-1}(U_N)^n=\exp(2iL\omega)\beta(L)
-i\frac{\epsilon_i}{N}\frac{1-(U_N)^N}{1-U_N}
\end{equation}
In the  $N\to \infty$ limit the denominator becomes $1-U_N\to -2i\omega L/N$ and 
\begin{equation}
\beta(L)=\exp(2i\omega L)\beta (-L)+\frac{\epsilon_i}{2\omega L}(1-e^{2i\omega L})\,.
\end{equation}
The first term is the free evolution of a neutral oscillator and the second term decays as $1/L$. The zero-frequency mode ($L\to\infty$) of the radiation gives no contribution to the time evolution of a charged oscillator. In other words, the momentum kick is averaged out.

\paragraph{Quantum treatment.}  Consider  Hamiltonian \eqref{Kamilton particle}  augmented by the harmonic oscillator potential term
\begin{equation}\label{Kamilton oscillator}
    \hat{K}=\frac{\hat{p}_i\hat{p}_i+\hat{p}_3^2}{2m}+\frac{1}{2}m\omega^2\hat{x}_a^2+q\hat{x}^{ i}\dot{\A}_i(\hat{x}).
\end{equation}
We will show below that if the quantum harmonic oscillator with Hamiltonian \eqref{Kamilton oscillator} is prepared in a energy eigenstate $|in\rangle=|\en\rangle$ at $u\to-\infty$, then its state at $u\to+\infty$ is given by 
\begin{equation}
    |out\rangle=\hat{S}|in\rangle\,,\qquad \hat{S}=1
\end{equation}
The free spectrum (when $q=0$) consists of states $|\en\rangle=|n_1,n_2,n_3\rangle$ with energy $E_\en=\omega(n_1+n_2+n_3+3/2)$ and $n_a\geq 0$. Let $|\en\rangle$ evolve with unperturbed Hamiltonian (in the interaction picture). The Schr\"odinger equation for an arbitrary state can be easily seen to take the following form
\begin{equation}
   i\frac{d}{du} \langle\en|\Psi\rangle=\sum_\mm\langle\en|q\hat{x}^{ i}\dot{\A}_i(\hat{x})|\mm\rangle e^{-iu(E_\en-E_\mm)}\langle \mm|\Psi\rangle\,.
\end{equation}
In the adiabatic approximation, where $\dot{\A}\ll \omega\A$, the quantity above can be nonvanishing only if $E_\en=E_\mm$. The position operator, however, has no diagonal element in energy eigen-state representation, $\langle\en|q\hat{x}^{ i}|\mm\rangle=0$ and $\langle\en|\Psi\rangle$ is constant in time. Therefore, the zero-frequency-mode of radiation has no effect on a harmonic oscillator, in contrast to the free particle.


\bibliographystyle{fullsort.bst}
 \providecommand{\href}[2]{#2}\begingroup\raggedright\endgroup


\end{document}